\documentclass{elsart}

\usepackage{epsfig}
\usepackage{amssymb}

\begin{document}

\sloppy

\begin{frontmatter}

\title{Excitation energy and strength of the pygmy dipole resonance in stable tin isotopes\thanksref{sponsor}}
\thanks[sponsor]{Work supported by the DFG under contract SFB 634.}

\author[a,b]{B. \"Ozel},
\author[a]{J. Enders},
\author[c]{H. Lenske},
\author[a,cor]{P. von Neumann-Cosel}\corauth[cor]{Corresponding author}\ead{vnc@ikp.tu-darmstadt.de},
\author[a]{I. Poltoratska},
\author[a]{V.Yu. Ponomarev},
\author[a,d]{A. Richter},
\author[a]{D. Savran},
\author[c]{N. Tsoneva}%,
%\author[a]{N. N.}%,
%\author[a]{A. Zilges},
%\author[a]{},

\address[a]{Institut f\"ur Kernphysik, Technische Universit\"at Darmstadt,
64289 Darmstadt, Germany}
\address[b]{Faculty of Science and Letter, \c{C}ukurova University,
01330 Adana, Turkey}
\address[c]{Institut f\"ur Theoretische Physik, Justus-Liebig-Universit\"at
Giessen,35392 Giessen, Germany}
\address[d]{ECT*, Villa Tambosi, I-38050 Villazano (Trento), Italy}

\begin{abstract}
The $^{112,120}$Sn$(\gamma,\gamma')$ reactions have been studied at
the S-DALINAC. Electric dipole (E1) strength distributions have been
determined including contributions from unresolved strength
extracted by a fluctuation analysis. Together with available data on
$^{116,124}$Sn, an experimental systematics of the pygmy dipole
resonance (PDR) in stable even-mass tin isotopes is established. The
PDR centroid excitation energies and summed strengths are in
reasonable agreement with quasiparticle-phonon model calculations
based on a nonrelativistic description of the mean field but
disagree with relativistic quasiparticle random-phase approximation
predictions.
\end{abstract}


\end{frontmatter}

% Introduction

The electric pygmy dipole resonance (PDR) in nuclei is a topic of
high current interest (for a recent review, see~\cite{paa07}). It is
expected to occur at energies well below the isovector giant dipole
resonance (IVGDR) and may exhaust a considerable fraction of the
total electric dipole strength in nuclei with a very asymmetric
proton-to-neutron ratio. Based on an analysis of transition
densities, most microscopic models qualitatively agree on its nature
as an oscillation of a neutron skin - emerging with an increasing
$N/Z$ ratio - against an approximately isospin-saturated core.
However, quantitative predictions of the centroid energy and
strength of the PDR as a function of neutron excess differ
considerably, in particular between models based on a relativistic
and a nonrelativistic description of the mean field, respectively.

While data in very neutron-rich heavy nuclei are scarce, the mode
has been investigated extensively utilizing the $(\gamma,\gamma')$
reaction in stable even-mass nuclides (see e.g.\
Refs.~\cite{kne06,end05} and references therein), in particular at
the shell closures $Z = 20$ \cite{har00}, $N = 50$ \cite{sch07}, $Z
= 50$ \cite{gov98}, $N = 82$ \cite{zil02,vol06,sav08} and in
$^{208}$Pb \cite{rye02}. Although the PDR is much weaker excited in
these nuclei, detailed spectroscopy provides important insight into
a possible interpretation of the mode as a neutron-skin oscillation,
the role of collectivity and single-particle degrees of freedom and
its isospin nature \cite{sav06}. However, the connection of these
results to the PDR in nuclei with very large $N/Z$ ratios still
remains a subject of debate \cite{paa07}.

In this respect, a systematic investigation of the PDR in the tin
isotope chain is of special interest. Recently, measurements of the
E1 response below the IVGDR in the exotic isotopes $^{130,132}$Sn
has been reported \cite{adr05}. Combined with results on the stable
isotopes, for the first time a set of data spanning a large range of
$N/Z$ ratios from 1.24 to 1.64 is thus available, which can serve as
a benchmark test for the validity of various theoretical approaches.
Indeed, the Sn isotopes have been a favorite case in the model
calculations to systematically investigate the features of the PDR
as a function of neutron excess
\cite{sar04,tso04a,vre04,paa05,kam06,pie06,ter06,tso08}.

Experimental information on the PDR in $^{116}$Sn and $^{124}$Sn is
available from Ref.~\cite{gov98}. Here we report results from new
$(\gamma,\gamma')$ experiments on $^{112}$Sn and $^{120}$Sn, which
allow the systematics of the PDR over the range of stable tin
isotopes to be established. These are compared to calculations
within the framework of the quasiparticle-phonon model (QPM) using a
nonrelativistic mean-field description and with the quasiparticle
random-phase approximation (QRPA) based on a relativistic mean-field
(RMF) description.

% Experiment

\begin{figure}[tbh]
\centerline {\epsfig{file=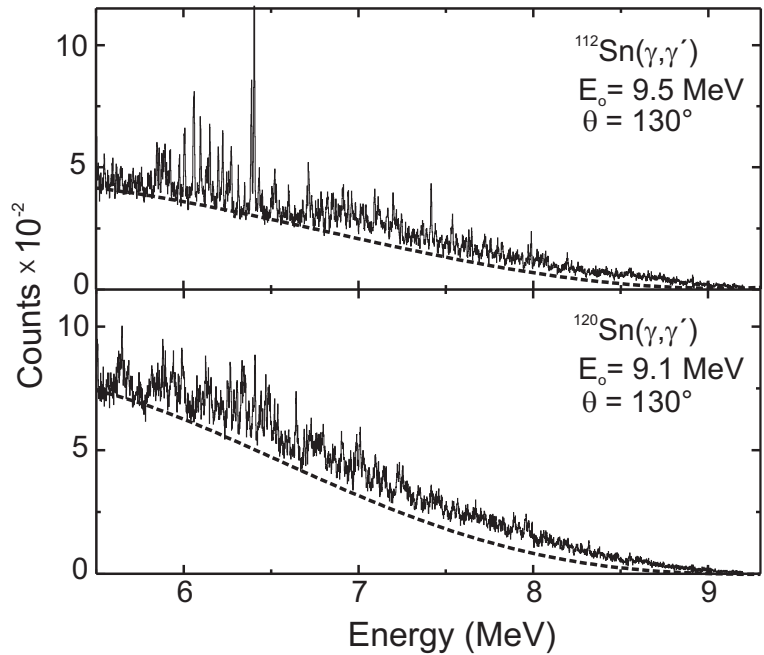,width=10cm}}
\caption{Spectra of the ($\gamma,\gamma'$) reaction at $\Theta =
130^\circ$ on $^{112}$Sn and $^{120}$Sn at $E_0 = 9.5$ MeV and 9.1
MeV, respectively, measured at the S-DALINAC. The dashed lines show
the nonresonant part of the spectrum deduced by a fluctuation
analysis described in the text based on $J^\pi = 1^-$ level
densities taken from the model of Ref.~\cite{rau97}.}
\label{fig:spec}
\end{figure}

The experiments on $^{112}$Sn and $^{120}$Sn were performed at the
superconducting Darmstadt electron linear accelerator S-DALINAC with
the nuclear resonance fluorescence (NRF) technique using electron
energies of 5.5, 7.0 and 9.5 MeV for $^{112}$Sn and 7.5 and 9.1 MeV
for $^{120}$Sn to generate bremsstrahlung. The maximum photon
energies were chosen below the neutron separation energies of both
nuclides to avoid the production of neutrons from ($\gamma$,n)
reactions, which would lead to a significant increase of the
background in the spectra. A detailed description of the
experimental setup can be found in Ref.~\cite{moh99} and
experimental details in Ref.~\cite{oze08a}. Targets consisted of
about 2 g highly enriched ($> 90\%$) $^{112}$Sn and $^{120}$Sn
sandwiched between two layers of boron. Well known transitions in
$^{11}$B were used to determine the photon spectrum and for the
energy calibration. Figure~\ref{fig:spec} presents
$(\gamma,\gamma^{\prime})$ spectra  for $ ^{112}$Sn at $E_0 = 9.5$
MeV (top) and  for $^{120}$Sn at $E_0 = 9.1$ MeV (bottom) taken at
an angle of ${130^{\circ}}$ with respect to the incident beam.
Significant differences are suggested by the data as the strength in
$^{120}$Sn is much more fragmented.

%B(E1) strength distribution

Reduced transition strengths were extracted for $^{112,120}$Sn as
explained e.g.\ in Ref.~\cite{kne06}. By analyzing the $\gamma$-ray
angular distributions the spins of all previously unknown states was
found to be $J = 1$; however, the parity was not determined. Thus
all dipole transitions were assumed to have E1 character based on
experimental findings in a large number of heavy semimagic nuclei
\cite{sch07,gov98,her97,pie02,bue08}. A branching ratio
$\Gamma_0/\Gamma = 1$ was assumed because no decay branch into
excited states was observed.
%Possible corrections for unobserved transitions to
%excited states were estimated to be $< 2$\% based on the QPM
%calculations described below.
Feeding effects \cite{kne06} were corrected for by utilizing the
comparison of results obtained at different endpoint energies.

\begin{figure}[tbh]
\centerline {\epsfig{file=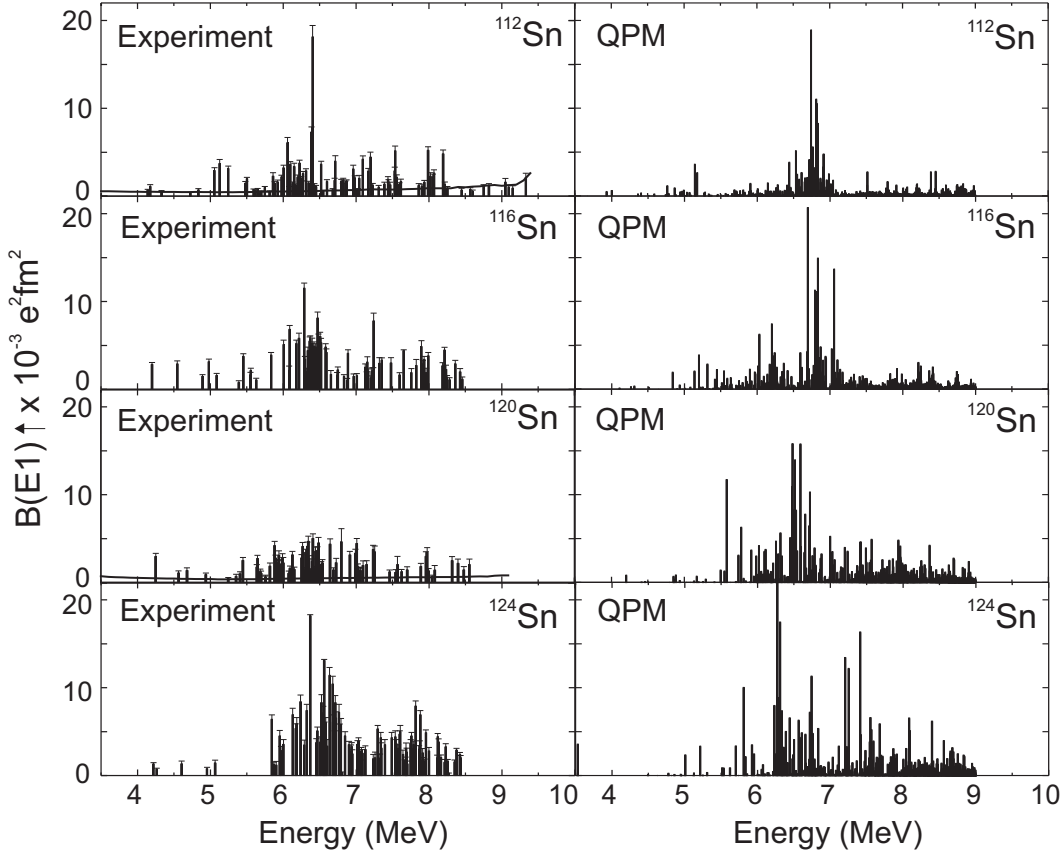,width=14cm}}
\caption{L.h.s.: Experimental B(E1) strength distributions in
$^{112,116,120,124}$Sn. The data for $^{112,120}$Sn are from the
present work, those for $^{116,124}$Sn from \cite{gov98}. R.h.s.:
QPM calculations of the corresponding B(E1) strength distributions
including up to three-phonon states described in the text. The
smooth lines indicate the sensitivity limits of the present
experiments.} \label{fig:be1}
\end{figure}

On the l.h.s.\ of Fig.~\ref{fig:be1} the B(E1) distributions
extracted in $^{112,120}$Sn are shown between 4 and 9 MeV and
compared to those in $^{116,124}$Sn measured previously by Govaert
et al.~\cite{gov98}. (Note that the prominent transitions resulting
from the population of the $[2^+ \otimes 3^-]_{1^-}$ two-phonon
states \cite{bry99,pys06} lie below 3.5 MeV and are therefore not
shown). All distributions exhibit a concentration of strength
between 6 and 7 MeV believed to represent the main part of the PDR.
However, there is also sizable strength at higher energies which
varies from isotope to isotope. Furthermore, a different
fragmentation pattern with smaller individual strengths in
$^{120}$Sn already indicated by the spectrum (Fig.~\ref{fig:spec})
is clearly visible.

The r.h.s.\ of Fig.~\ref{fig:be1} shows results from calculations
with the QPM, where the mean field is taken from a global
parametrization \cite{pon79} and levels near the Fermi surface are
adjusted to experimental values. One-, two- and three-phonon states
were included similar to Ref.~\cite{sav08}, which should give a
rather complete spectrum of $1^-$ states up to about 7 MeV. As in
previous calculations within this scheme
\cite{gov98,sav08,rye02,her97}, the transition densities of states
forming the PDR exhibit the features of a neutron-skin oscillation
against an isospin-saturated core. Figure~\ref{fig:be1} demonstrates
that good correspondence between the experimentally observed and
calculated fine structure can be obtained when the coupling to
complex configurations, i.e.\ up to three-phonon states,  is taken
into account.

%Fluctuation analysis

The sensitivity limits of the present experiment for detecting a
$\gamma$-ray transition are indicated in Fig.~\ref{fig:be1} by the
smooth lines. A comparison to the QPM calculations indicates that
sizable strength could be missing. Furthermore, unresolved strength
due to the finite energy resolution may be hidden in the spectra. A
statistical analysis of the PDR in $N = 82$ nuclei \cite{end04}
suggested that the level density in the region of the PDR is very
high. Current level-density models based on the backshifted Fermi
gas (BSFG) approach \cite{rau97,egi05} and microscopic HF-BCS
calculations \cite{dem01} predict average level spacings (the
inverse of the level density $\rho$) in the energy range of interest
up to $1/\rho \approx 1$ keV in the investigated tin isotopes, i.e.\
values below the typical Ge detector resolution $\Delta E = 4 - 7$
keV in the $\gamma$-energy range of the present experiment. Thus
some of the E1 strength is indeed expected to lie in the background
of the measured spectra. Such unresolved strength can be extracted
by means of a fluctuation analysis \cite{han90}. Application of this
technique to $(\gamma,\gamma^{\prime})$ spectra is discussed in
detail in Refs.~\cite{end97,hux99,nor03}. The analysis requires
either a knowledge of the $1^-$ level density or of the nonresonant
background contributions to the spectra. Since the approach for a
model-independent determination of the nonresonant background based
on a wavelet decomposition \cite{kal06,kal07} does not work in the
case of the $(\gamma,\gamma')$ spectra\footnote{Application of the
technique requires a compact resonance signal and a sufficient
signal-to-background ratio in the spectrum.}, level densities were
taken from the three models discussed above . All three models
\cite{rau97,egi05,dem01} yield roughly the same magnitude and slope
of the nonresonant background in the $^{112,120}$Sn spectra. Results
obtained with the level density predictions of Ref.~\cite{rau97} are
shown as examples. Their actual slight differences
(indistinguishable within the line thickness for the case of
$^{112}$Sn) allow to estimate the systematic uncertainties caused by
the level density models in the extraction of the hidden E1
strength.

The method is applicable in the excitation region $E_{\rm x} = 5.5 -
7.8$ MeV. At lower energies the individual levels do not overlap
while at higher energies uncertainties in the present method get too
large because the fluctuation signal is too small.  A comparison of
the analysis of discrete transitions to the present one shows an
increase of the total E1 strength of 44(6)\% in $^{112}$Sn and
47(12)\% in $^{120}$Sn, respectively, due to unresolved levels when
integrating up to $E_{\rm x} = 7.8$ MeV. The quoted uncertainties
include the model dependence of the level densities and
finite-range-of-data errors.

%Systematics

\begin{figure}[tbh]
\centerline {\epsfig{file=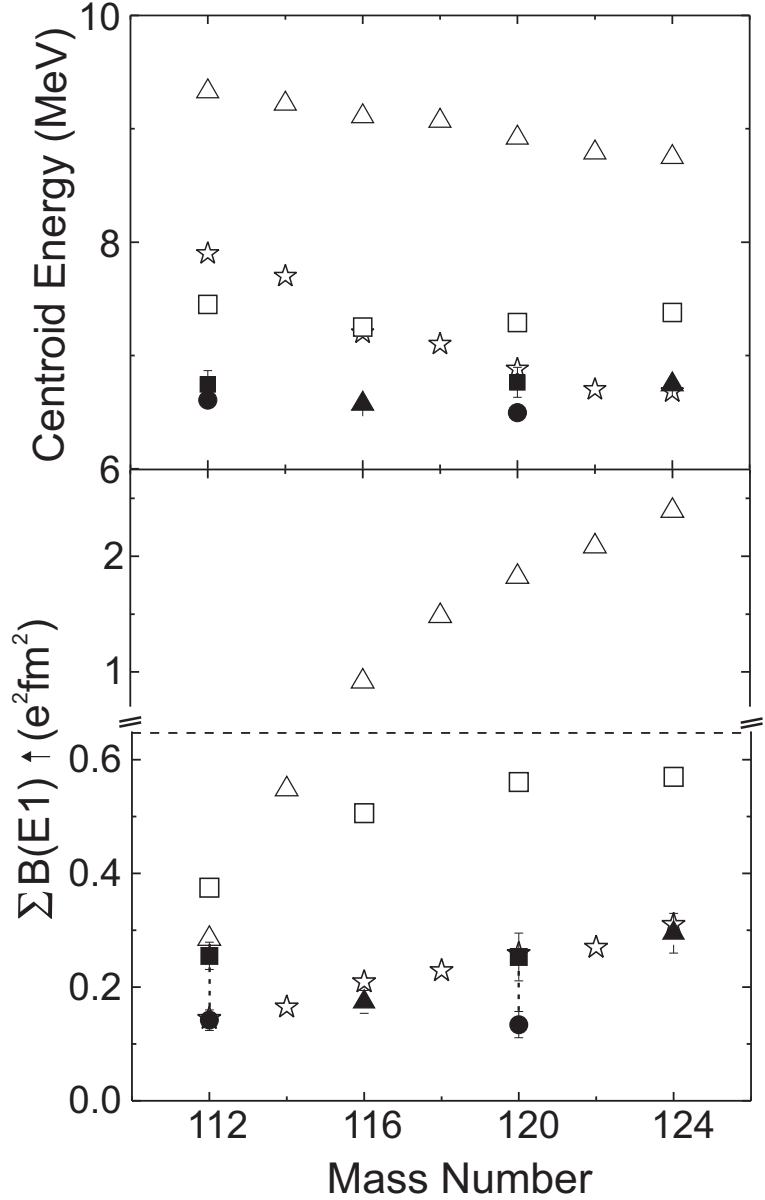,width=10cm}}
\caption{Systematics of the energy centroids (top) and summed E1
strengths (bottom) of the PDR in the Sn isotope chain. Experimental
results are shown as full triangles ($^{116,124}$Sn,
Ref.~\cite{gov98}), circles ( $^{112,120}$Sn, present experiment,
discrete transitions only) and squares ($^{112,120}$Sn, unresolved
strength included). Model results are shown as open stars (QPM,
Ref.~\cite{tso08}), open squares (QPM described in the text) and
open triangles (RQRPA, Ref.~\cite{paa05}).} \label{fig:sys}
\end{figure}

In Fig.~\ref{fig:sys} the systematics of the E1 strength in stable
tin isotopes summed in the energy interval $ 5 \leq E_{\rm x} < 9$
MeV is displayed as a function of the mass number. Two values are
shown for $^{112,120}$Sn investigated in the present work: full
circles represent the sum of strength from discrete transitions only
(which makes the result comparable to the $^{116,124}$Sn data from
\cite{gov98}) while full squares include the unresolved strength
just discussed.

The results are confronted with model predictions of the PDR
centroid energies (top) and summed strengths (bottom). Starting with
a discussion of the centroids, the experimental results (full
circles and triangles) appear to be nearly constant at a value
$E_{\rm x} \approx 6.5$ MeV independent of the neutron excess.
Inclusion of the unresolved strengths obtained in $^{112,120}$Sn
(full squares) leaves the results almost unaffected. Such a behavior
is reproduced by the QPM calculation introduced above (open
squares), albeit the centroids lie systematically about 500 keV
higher. The QPM approach of Ref.~\cite{tso08} (open stars) predicts
a systematic dependence on the neutron excess. Good correspondence
with the data is observed for the heavier $^{120,124}$Sn isotopes
but for $^{112,116}$Sn the experimental centroids are significantly
lower. The predictions of a self-consistent relativistic QRPA
approach based on the relativistic Hartree-Bogoliubov model, using
an interaction with density-dependent meson-nucleon couplings
(DD-ME2) \cite{paa05}, are presented as open triangles. They show a
weak mass dependence but the predicted centroids are about 2 MeV
higher than the data.

The experimental summed E1 strengths (bottom part of
Fig.~\ref{fig:sys}) display the following pattern: Analyzing the
discrete transitions only, an increase with increasing neutron
number is observed - as predicted by all calculations shown here -
but the E1 strength shows a minimum for $^{120}$Sn. Including
unresolved transitions raises the strength by about 50\% for
$^{112,120}$Sn. Its magnitude depends sensitively on the $1^-$ level
densities which are expected to vary little (typically less than a
factor of two) over the range of investigated tin isotopes. Thus,
the contribution from unresolved transitions in $^{116,124}$Sn
should be similar leaving the pattern unchanged. Comparing to the
data including the unresolved strength, the QPM results of
Ref.~\cite{tso08} are somewhat below the data for $^{112}$Sn (and
probably also for $^{116,124}$Sn) and agree for $^{120}$Sn. The QPM
calculations described above give PDR strengths which are typically
a factor of two larger than experiment. Finally, the relativistic
QRPA (open trinagles) dramatically overpredicts the strength (note
the scale change in Fig.~\ref{fig:sys}) except for $^{112}$Sn.  A
similar RQRPA calculation with a slightly different interaction has
been reported by Piekarewicz \cite{pie06} with almost identical
results for the PDR centroids and strengths.

We note that the excitation energy intervals used for the summation
in the model calculations partly differ from the experimental one
since the PDR strengths are predicted in different excitation energy
regions (cf.\ the top part of Fig.~\ref{fig:sys}). In
Refs.~\cite{tso08,paa05} the intervals are determined based on an
analysis of the transition densities. For the QPM calculations
presented above, the summation is performed over the energy range of
the data. The QPM results of Ref.~\cite{tso08} correspond to an
energy range $E_{\rm x} < 8.1$ MeV (based on the structure
argument). If one would sum up to 9 MeV as in the experimental case,
the E1 strengths would be significantly larger (cf.\ Fig.~7 in
\cite{tso08}). The quoted PDR properties from the RQRPA calculations
correspond to a summation up to 10 MeV.

Clearly, none of the models provides a satisfactory reproduction of
both quantities, viz.\ energy centroid and strength of the PDR, in
the stable tin isotopes. However, the QPM calculations are generally
much closer to the data. The variations between neighboring nuclei
and the small total E1 strengths of less than 1\% of the EWSR
suggest that the transitions retain much of their single-particle
character. Oros et al.~\cite{oro98} demonstrated for example in a
schematic two-group RPA calculation that irregularities of the
unperturbed one-particle one-hole (1ph) spectrum can lead to a local
concentration of E1 strength well below the IVGDR. On the other
hand, it is instructive to recall the important role of mixing
between transitions considered to belong to the PDR and slightly
higher-lying one-phonon states which lead to a shift of the
low-energy single-particle continuum below the threshold (see e.g.\
Ref.~\cite{tso04b}). This mechanism is also very sensitive to
details of the mean-field description.

% Summary

To summarize, we have presented a NRF study of the dipole strength
in $^{112}$Sn and $^{120}$Sn up to the neutron threshold. The
analysis reveals in both nuclei significant unresolved strength in
the energy region $E_x = 5.5 - 7.8$ MeV which must be included when
extracting integral features of the dipole strength distribution
from the data. Combined with previous results on $^{116,124}$Sn from
Ref.~\cite{gov98}, for the first time a systematics of the PDR in
the stable tin isotopes is now available. Their global features are
in reasonable agreement QPM calculations but disagree with RQRPA
results which predict significantly higher centroid energies and
larger collectivity. An extrapolation of the these results to the
exotic $^{130,132}$Sn isotopes suggests that the low-energy E1
strength observed in the Coulomb breakup experiments \cite{adr05}
does not represent the PDR, but rather some other nuclear effect.

The collectivity of the PDR in the stable tin isotopes predicted by
the models discussed above differs substantially ranging from an
interpretation as almost pure 1ph transition with a $\nu[3p_{3/2}
3s_{1/2}^{-1}]$ structure \cite{tso08} to a rather collective mode
exhausting up to about 7\% of the EWSR \cite{paa05,pie06}. Very
recently, for the first time RQRPA calculations of the E1 strength
in tin isotopes including phonon coupling have been presented
\cite{lit07,lit08}. While the global characteristics of the PDR
remain essentially unchanged, due to the fragmentation some strength
is shifted into the energy region below 7 MeV which is claimed to be
in reasonable accordance with the strengths experimentally observed
in $^{116,124}$Sn \cite{gov98}. Still, the discrepancy with
predictions of the QPM remains since the latter finds the
characteristic features of the PDR (based on the structure of the
transition densities \cite{rye02,vre01}) for states below $E_{\rm x}
\approx 7$ MeV only. Calculations of the fine structure of the E1
mode in the various models are called for, such as the ones shown in
Fig.~\ref{fig:be1}. This should allow to further elucidate the
structure of the PDR.

Uncertainties on the experimental side remain also because of the
inherent limitations of the used techniques. The Coulomb breakup
experiments are restricted to energies above neutron threshold and
the NRF experiments (roughly) to energies below threshold.
Furthermore, some correction of the NRF results may be necessary due
to unobserved decay to excited states. New experimental approaches
are thus desired to determine complete E1 strength distributions
from low energies up to the GDR region with good resolution.
Therefore, a tagger for this energy region aiming at a resolution
$\Delta E / E = 25$ keV (FWHM) is presently put into operation at
the S-DALINAC \cite{lin07}. Alternatively, intermediate-energy
polarized proton scattering at $0^\circ$ has been developed as a new
spectroscopic tool for dipole strength with comparable energy
resolution \cite{tam07}. As demonstrated for the case of $^{208}$Pb,
E1 strengths extracted from such data are in good agreement with NRF
studies \cite{vnc08a}. Because of the conflicting results in the
present work a case study of $^{120}$Sn is presently underway
\cite{vnc08b}.

\section*{Acknowledgements}
Discussions with G.~Col\`{o}, N.~Paar and D.~Vretenar are
acknowledged. We are indebted to N.~Paar and J.~Piekarewicz for
providing us with numerical results of their calculations and to
Y.~Kalmykov for his help with the fluctuation analysis. We thank the
GSI for the loan of the enriched $^{112}$Sn target. B\"O
acknowledges financial support by the DAAD sandwich program.

\end{document}